\documentclass[aps,prd,twocolumn,showpacs,nofootinbib]{revtex4-1}

\usepackage{amsfonts,color,graphicx,epsfig,amsmath,multirow,slashed}

\begin{document}

\title{$\mathcal O(\alpha^{3}\alpha_s)$ Study on the yields and polarizations of $J/\psi(\Upsilon)$ within the framework \\
of non-relativistic QCD via $\gamma\gamma \to J/\psi(\Upsilon)+\gamma+X$ at CEPC}
\author{Zhan Sun}
\email{zhansun@cqu.edu.cn (corresponding author)}
\address{Department of Physics, Guizhou Minzu University, Guiyang 550025, P.R. China}

\author{Xiong Deng}
\address{Department of Physics, Guizhou Minzu University, Guiyang 550025, P.R. China}

\author{Hai Bing Fu}
\email{fuhb@cqu.edu.cn}
\address{Department of Physics, Guizhou Minzu University, Guiyang 550025, P.R. China}

\date{\today}

\begin{abstract}
Within the framework of the non-relativistic QCD (NRQCD), we make a systematical study of the yields and polarizations of $J/\psi$ and $\Upsilon$ via $\gamma \gamma \to J/\psi(\Upsilon)+\gamma+X$ in photon-photon collisions at the Circular Electron Positron Collider (CEPC), up to $\mathcal O(\alpha^{3}\alpha_s)$. We find that this process at CEPC is quite ``clean", namely the direct photoproduction absolutely dominate over the single- and double- resolved processes, at least 2 orders of magnitude larger. It is found that the next-to-leading order (NLO) QCD corrections will significantly reduce the results due to that the virtual corrections to $^3S_1^1$ is large and negative. For $J/\psi$, as $p_t$ increases, the color octet (CO) processes will provide increasingly important contributions to the total NLO results. Moreover the inclusion of CO contributions will dramatically change the polarizations of $J/\psi$ from toally transverse to longitudinal, which can be regarded as a distinct signal for the CO mechanism. However, for the case of $\Upsilon$, the effects of the CO processes are negligible, both for yields and polarizations. For $J/\psi$, the dependence of the yields on the value of the renormalization scale $\mu_r$ is moderate, while significant for the polarization. The impact of the variation of $\mu_{\lambda}$ is found to be relatively slight. As for the case of $\Upsilon$, the uncertainties of $\mu_{r}$ and $\mu_{\lambda}$ just bring about negligible effects. The future measurements on this semi-inclusive photoproductions of $J/\psi(\Upsilon)+\gamma+X$, especially on the polarization parameters of $J/\psi$, will be a good laboratory for the study of heavy quarkonium production mechanism and helpful to clarify the problems of the $J/\psi$ polarization puzzle.
\pacs{13.66.Bc, 12.38.Bx, 12.39.Jh, 14.40.Pq}

\end{abstract}

\maketitle

\section{Introduction}
The non-relativistic QCD (NRQCD) factorization \cite{Bodwin:1994jh} is one of the most effective method to describe the heavy quarkonium processes, which provides a systematical way to separate the effects from the dynamics at different scales. That is, the production process can be factorized into a sum of products of short-distance coefficients (SDCs) and the long distance matrix elements (LDMEs). The SDCs are perturbatively calculable in a power series of $\alpha_s$ and the LDMEs can be extracted by fitting the predictions with the experimental data. In many processes, especially the hadroproductions, NRQCD are quite successful, such as the NRQCD predictions filled the large gap between the leading order (LO) color singlet (CS) predictions and the measurements of $J/\psi$ and $\psi(2S)$ hadroproductions released by CDF group \cite{Braaten:1994vv,Cho:1995ce,Cho:1995vh}. In addition, the measurements of $\chi_c$ and $\eta_c$ hadroproductions can also be described well within NRQCD \cite{Ma:2010vd,Jia:2014jfa,Han:2014jya,Zhang:2014ybe}. Moreover, recent works on the $J/\psi$ electroproductions in deeply inelastic $ep$ scattering (DIS) at HERA reveal that the CS contributions are not impossible to saturate the measurements and the inclusion of the color octet (CO) partonic processes will make the agreement between theory and experiment much better \cite{Sun:2017nly,Sun:2017wxk}. Although the NRQCD factorization has gained its reputation from the success in many processes, it is still facing several challenges. Such as the NRQCD results for $e^+e^- \to J/\psi+X_{\textrm{non}-c\bar{c}}$ severely overshoot the experimental measurements \cite{Zhang:2009ym}, which is recognized as one of the most serious challenges on the universality of the CO LDMEs. In addition, the long standing $J/\psi$ polarization puzzle is another controversy for NRQCD. Different sets of LDMEs will result in totally different predicted polarization parameters \cite{Butenschoen:2012px,Chao:2012iv,Gong:2012ug,Shao:2012fs,Shao:2014fca,Shao:2012fs,Bodwin:2014gia,Sun:2015pia}. That is to say, the theoretical predictions based on NRQCD , especially the polarization parameters, severely rely on the values of these LDMEs. Therefore, it is indispensable to further test the validity of NRQCD in various kinds of experiments.

The semi-inclusive photoproduction of $J/\psi(\Upsilon)+\gamma+X$ at CEPC can be a good candidate for identifying the significance of the CO mechanism. On one hand, taking $J/\psi$ for example, when the transverse momentum of $J/\psi$ is not too large, the direct photoproduction $\gamma\gamma \to J/\psi+\gamma+X$ is found to be dominant, at least two orders of magnitude bigger than the single- ($\gamma g \to c\bar{c}[^3S_1^8]+\gamma+X$) and double- ($gg \to c\bar{c}[^3S_1^1,^1S_0^8,^3S_1^8,^3P_J^8]+\gamma+X,q\bar{q} \to c\bar{c}[^3S_1^8]+\gamma+X$) resolved contributions which are suppressed strongly by the suppression of the CO LDMEs and/or the gluon (quark) PDF of the photon\footnote{For the photon density function of Weizsacker-Williams approximation (WWA), describing the initial bremsstrahlung photons which are generated in a circular $e^+e^-$ collider, the suppression via PDF is significant. However, for the Laser Back Scattering (LBS) photon generated in a high energy Liner $e^+e^-$ Collider, the density function is far different from WWA and the single- and double- resolved gluon subprocess can be significant at low $p_t$ region, even dominant over the direct one as is presented in Figure \ref{fig:ptJpsiresolve}, analogous to that in Ref \cite{Li:2009fd}. Due to CEPC \cite{CEPC1,CEPC2} is a circular electron positron collider, in our present paper, only WWA photon is considered.}, as is demonstrated in Figure \ref{fig:ptJpsiresolve}. Note that, for the gluon-fragmentation process $q\bar{q} \to c\bar{c}[^3S_1^{8}]+\gamma$, only at sufficiently large $p_t$ regions, the kinematic enhancements due to the scales $p_t^{-4}$ can compensate for the suppressions of CO LDMEs and PDFs, even be dominant over the direct case. However, the events therein are several orders of magnitude smaller than that at low and mid $p_t$ regions. In conclusion, the process $\gamma\gamma \to J/\psi(\Upsilon)+\gamma+X$ is very ``clean". On the other hand, since the resolved processes can be safely neglected, we do not have to take the gluon saturation effects \cite{Mueller:1985wy} into consideration. In addition, for $\gamma\gamma \to J/\psi(\Upsilon)+\gamma+X$, up to $\mathcal O(\alpha^{3}\alpha_s)$, only two CO processes are involved, namely $^3P_J^8$ and $^1S_0^8$, and the $^3S_1^8$ partonic process is forbidden due to the conservation of $C-$parity or color. The one LDME short makes the extraction of the LDMEs easier than that at LHC. In view of these advantageous features, the photoproductions of $J/\psi(\Upsilon)+\gamma+X$ at CEPC are quite suitable for studying the production mechanism of the heavy quarkonium.

There have been several studies on $J/\psi+\gamma+X$ in $pp,~ep$ and $\gamma\gamma$ collisions \cite{Drees:1991ig,Doncheski:1993dj,Kim:1994bm,Mirkes:1994jr,Roy:1994vb,Kim:1996bb,Mathews:1999ye,Klasen:2004az,Kniehl:2002wd,Li:2008ym,Li:2014ava}. In the present work, we will systematically study the semi-inclusive photoproduction of $J/\psi+\gamma+X$ in photon-photon collisions at CEPC, for the first time investigating the polarisation parameters for $\gamma\gamma \to J/\psi+\gamma+X$, which is proved to be a distinct signal for testing the CO mechanism. As stated before, the NRQCD predictions always heavily suffer from the uncertainties of the LDMEs, thus for the production of $J/\psi$ we will adopt two different sets of LDMEs on the market to present our predictions. Comparing with charmonium, $b\bar{b}$ mesons have their own ideal properties. Due to $b\bar{b}$ is much heavier than $c\bar{c}$, both the typical coupling constant and relative velocity between the two heavy flavor constituent quarks are smaller than those of charmonium. Therefore, the perturbative results for $b\bar{b}$ may be more convergent over the expansion of $\alpha_s$ and $v^2$ than charmonium. Moreover, the dependance on $\mu_r$ is expected to be more moderate. Considering these advantages, along the same lines, for the first time we will make a complete study of the yields and polarizations of $\Upsilon$ via $\gamma\gamma \to \Upsilon+\gamma+X$ in photon-photon collisions to provide a sound estimate.

The remaining parts of the paper are organized as follows. In Sec.II, we describe the calculation technology and introduce the input parameters employed in our calculations. In Sec.III, we present the phenomenological results, including $p_t$ distributions and polarization parameter $\lambda$. The last section is reserved for a summary.
\section{Calculation Formalism}
Within the NRQCD framework, the differential cross section for the process $e^{+}e^{-} \rightarrow e^{+}e^{-}+H(Q\bar{Q})+\gamma+X$ can be factorized as
\begin{eqnarray}
d\sigma&=&\sum_{n}\int dx_1 dx_2 f_{\gamma}(x_1)f_{\gamma}(x_2)\sum_{i,j}\int dx_i dx_j f_{i/\gamma}(x_i) \nonumber \\
&&f_{j/\gamma}(x_j){d\hat{\sigma}(i+j\to Q\bar{Q}[n]+\gamma+X)}\times\langle \mathcal O ^{H}(n) \rangle \nonumber \\
\end{eqnarray}
where $\hat{\sigma}$ is the parton-level short-distance coefficients (SDCs) representing the production of a configuration of the $Q\bar{Q}$ intermediate state with the quantum number $n$ and $\langle \mathcal O ^{H}(n) \rangle$ is the nonperturbative but universal LDMEs. $f_{\gamma}(x)$ is the photon density function and $f_{i/\gamma}(x_i)$ is the parton distribution function of the photon. Here $x$ and $x_{i,j}$ represent the momentum fraction of the initial photon to the initial electron or positron and the hadronic contents ($i,j$), such as gluon and light quarks, to the initial photon, respectively. At CEPC, the initial photons are mainly generated by bremsstrahlung, for which the spectrum can be described by the Weizsacker-Williams approximation (WWA) \cite{Frixione:1993yw} as:
\begin{eqnarray}
f_{\gamma}(x)&=&\frac{\alpha}{2\pi}[2m^2_e(\frac{1}{Q^2_{max}}-\frac{1}{Q^2_{min}})x \nonumber \\
&+&\frac{1+(1-x)^2}{x}\log(\frac{Q^2_{max}}{Q^2_{min}})]
\end{eqnarray}
with
\begin{eqnarray}
&&Q^2_{min}=\frac{m^2_e x^2}{1-x}, \nonumber \\
&&Q^2_{max}=(\frac{\sqrt{s}\theta}{2})^2(1-x)+Q^2_{min}.
\end{eqnarray}
where $x=\frac{E_{\gamma}}{E_e}$ denotes the ratio of the electron (positron) momentum taken by the interacting photon, $\alpha$ is the fine structure constant and $m_e$ represents the electron mass. $\theta$ is the angle between the flying direction of the initial photon and the direction of the electron (positron) beam. In our present work, the $\theta$ is taken as $32$mrad as at LEPII.

As stated before, for $\gamma\gamma \to J/\psi(\Upsilon)+\gamma+X$ at CEPC, the contributions from single- and double- resolved photon processes are negligible, hence our QCD corrections only focus on the direct photoproduction. Up to $\mathcal O(\alpha^{3}\alpha_s)$, for $\gamma\gamma \to J/\psi(\Upsilon)+\gamma+X$, there are one virtual correction and two real correction processes, namely
\begin{eqnarray}
\gamma+\gamma &\to& J/\psi(\Upsilon)[^3S_1^1]+\gamma, \\ \label{3s11}
\gamma+\gamma &\to& J/\psi(\Upsilon)[^1S_0^8]+\gamma+g, \\ \label{1s08}
\gamma+\gamma &\to& J/\psi(\Upsilon)[^3P_J^8]+\gamma+g. \label{3pj8}
\end{eqnarray}
There are 78 Feynman diagrams for the virtual correction, among which 48 for one-loop and 30 for counter-term. As for the two real processes, there are 24 diagrams for each channel. The dimensional renormalization with $D=4-2\epsilon$ is adopted to isolate the ultraviolet (UV) and infrared (IR) singularities and the Coulomb singularities are absorbed into the redefined LDMEs. The on-mass-shell (OS) scheme is adopted to set the renormalization constants of the quark mass $Z_m$ and the field $Z_2$ as following:
\begin{eqnarray}
\delta Z^{OS}_m&=&-3 C_F  \frac{\alpha_s}{4\pi}[\frac{1}{\epsilon_{UV}}-\gamma_E+\ln\frac{4\pi \mu^2}{m^2}+\frac{4}{3}], \\
\delta Z^{OS}_2&=&- C_F  \frac{\alpha_s}{4\pi}[\frac{1}{\epsilon_{UV}}+\frac{2}{\epsilon_{IR}}-3\gamma_E+3\ln\frac{4\pi \mu^2}{m^2}+4]
\nonumber \\
\end{eqnarray}
where $C_F=\frac{4}{3}$ and $\gamma_E$ reprensents the Euler's constant.

The cross section for Eq.(\ref{3pj8}) also has IR singularity, which can be eliminated by calculating the NLO corrections to the LDMEs $\langle \mathcal O[^3S_1^8] \rangle$ and $\langle \mathcal O[^3S_1^1] \rangle$. Taking $\langle \mathcal O^{J/\psi}[^3S_1^1] \rangle$ for example, the bare LDME can be written as
\begin{eqnarray}
\langle \mathcal O^{J/\psi}[^3S_1^1] \rangle_{\textrm{bare}}&=&\langle \mathcal O^{J/\psi}[^3S_1^1] \rangle-\frac{4\alpha_s}{3 \pi m^2_c} \nonumber \\
&\times&(\frac{1}{\epsilon_{IR}}-\frac{1}{\epsilon_{UV}})\langle \mathcal O^{J/\psi}[^3P_0^8] \rangle, \label{3pj81}
\end{eqnarray}
where $N_c=3$ for SUN(3) gauge theory and $m_c$ denotes the mass of $c-$quark. The $\mu_{\lambda}-$cutoff renormalization scheme is adopted to subtract the UV divergence. By substituting the relation between the bare and renormalized LDMEs,
\begin{eqnarray}
&&\langle \mathcal O^{J/\psi}[^3S_1^1] \rangle_{\textrm{bare}} \nonumber \\
&=&\langle \mathcal O^{J/\psi}[^3S_1^1] \rangle_{\textrm{renorm}}+\frac{4\alpha_s}{3 \pi m^2_c} \nonumber \\
&\times&\left[\frac{1}{\epsilon_{UV}}-\gamma_{E}+\frac{5}{3}+\ln(\frac{\pi \mu_r^2}{\mu_{\lambda}})\right]\langle \mathcal O^{J/\psi}[^3P_0^8] \rangle,
\end{eqnarray}
into Eq. (\ref{3pj81}), we have
\begin{eqnarray}
&&\langle \mathcal O^{J/\psi}[^3S_1^1] \rangle_{\textrm{renorm}} \nonumber \\
&=&\langle \mathcal O^{J/\psi}[^3S_1^1] \rangle-\frac{4\alpha_s}{3 \pi m^2_c} \nonumber \\
&\times&\left[\frac{1}{\epsilon_{IR}}-\gamma_{E}+\frac{5}{3}+\ln(\frac{\pi \mu_r^2}{\mu_{\lambda}})\right]\langle\mathcal O^{J/\psi}[^3P_0^8] \rangle,
\end{eqnarray}
Thus the process $\gamma\gamma \to c\bar{c}[^3P_J^8]+\gamma+g$ can be redefined as
\begin{eqnarray}
&&d\hat{\sigma}_{\textrm{renorm}}(\gamma\gamma \to c\bar{c}[^3P_J^8]+\gamma+g) \nonumber \\
&=&d\hat{\sigma}(\gamma\gamma \to c\bar{c}[^3P_J^8]+\gamma+g) \nonumber \\
&-&\frac{4\alpha_s}{3 \pi m^2_c}\left[\frac{1}{\epsilon_{IR}}-\gamma_{E}+\frac{5}{3}+\ln(\frac{\pi \mu_r^2}{\mu_{\lambda}})\right] \nonumber \\
&\times&d\hat{\sigma}(\gamma\gamma \to c\bar{c}[^3S_1^1]+\gamma). \label{softhard}
\end{eqnarray}
The soft parts of $d\hat{\sigma}(\gamma\gamma \to c\bar{c}[^3P_J^8]+\gamma+g)$, where the final gluon attached to $c\bar{c}[^3P_J^8]$ is soft, incorporate the terms relative to $d\hat{\sigma}(\gamma\gamma \to c\bar{c}[^3S_1^1]+\gamma)$ as\footnote{Notice that, the partonic process $\gamma\gamma \to c\bar{c}[^3S_1^8]+\gamma$ is forbidden, thus in our calculations the soft part of the cross section of $^3P_J^8$ does not contain the $^3S_1^8$ content.}
\begin{eqnarray}
&&\frac{4\alpha_s}{3 \pi m^2_c}\left[\frac{1}{\epsilon_{IR}}+\frac{p_0}{|\textbf{p}|}\ln(\frac{p_0+|\textbf{p}|}{p_0-|\textbf{p}|})+\ln(\frac{4 \pi \mu^2_r}{s \delta_s^2})-\gamma_{E}-\frac{1}{3}\right] \nonumber \\
&\times&d\hat{\sigma}(\gamma\gamma \to c\bar{c}[^3S_1^1]+\gamma)
\end{eqnarray}
where $p_0$ and $\textbf{p}$ are the energy and 3-momentum of $J/\psi$. Eventually, the cross section of $^3P_J^8$ is finite. Taking advantage of the relations $\langle \mathcal O^{J/\psi}(^3P_2^{[8]})\rangle=5\langle \mathcal O^{J/\psi}(^3P_0^{[8]})\rangle$, and $\langle \mathcal O^{J/\psi}(^3P_1^{[8]})\rangle=3\langle \mathcal O^{J/\psi}(^3P_0^{[8]})\rangle$, we can synthesise the three SDCs for $n=^3P_0^{[8]}$, $n=^3P_1^{[8]}$, and $n=^3P_2^{[8]}$ by defining
\begin{eqnarray}
&&d\hat{\sigma}(\gamma\gamma \to c\bar{c}[^3P_J^8]+\gamma+g) \nonumber\\
&=&d\hat{\sigma}(\gamma\gamma \to c\bar{c}[^3P_0^8]+\gamma+g) \nonumber \\
&+&3d\hat{\sigma}(\gamma\gamma \to c\bar{c}[^3P_1^8]+\gamma+g) \nonumber \\
&+&5d\hat{\sigma}(\gamma\gamma \to c\bar{c}[^3P_2^8]+\gamma+g) \nonumber
\end{eqnarray}
In addition to the yields, we will make a discussion on the polarization of $J/\psi(\Upsilon)$ in the $\gamma$ associated photoproductions. The polarization parameter $\lambda$ within the helicity frame is defined as \cite{Beneke:1998re}:
\begin{eqnarray}
\lambda=\frac{d\sigma_{11}-d\sigma_{00}}{d\sigma_{11}+d\sigma_{00}}
\end{eqnarray}
where $d\sigma_{S_z,S^{'}_z}(S_z,S^{'}_z=0,\pm1)$ denote the spin density matrix elements.

In this paper, we adopt the $FDC$ package \cite{Wang:2004du} to complete all the calculations. As a cross check, the virtual correction are as well accomplished by means of a highly automatic mathematica-fortran package-$Malt@FDC$.
\section{Numerical results and discussions}

\subsection{Input parameters}

During the calculations, we take $M_{J/\psi}=2m_c$ and $M_{\Upsilon}=2m_b$ with $m_c=1.5$GeV and $m_b=4.9$GeV. The fine structure constant $\alpha=\frac{1}{137}$. Since the LO partonic process is a pure QED process, for the NLO calculations the one-loop $\alpha_s$ running is adopted with $\alpha_s(M_Z)=0.13$. The default value of renormalization scale $\mu_r$ is set to be $\mu_0=m_T=\sqrt{p_t^2+4m^2_{c,b}}$ and $\mu_{\lambda}$ is taken as $m_{c,b}$. The center-of-mass energy at CEPC is 240 GeV. We adopt two sets of LDMEs on the market for $J/\psi$ production and one set for $\Upsilon$\footnote{In our calculations, it is found that the CO contributions to the production of $\Upsilon$ are quite tiny, almost negligible, thus we will only take one set of LDMEs for the case of $\Upsilon$.}, the detailed values of which can be found in Table \ref{LDME}, to present our numerical results. For convenience, hereinafter we name the LDMEs of \cite{Zhang:2014ybe,Sun:2015pia} and \cite{Bodwin:2014gia} set 1 and set 2 respectively. Notice that the initial bremsstrahlung photons may directly serve as the final state photon, in order to exclude these events, we adopt the criterion $p^{\gamma}_t>p^{\gamma}_{t,min}$ with $p^{\gamma}_{t,min}=1,2,3$ GeV, respectively. Furthermore, in our concerned semi-inclusive processes, together with $J/\psi(\Upsilon)$, the final state photons will be as well observed, therefore, for the experimental aspects, cuts on the transverse momentum of the final photon is indispensable. Considering the perturbative expansion may do not work well in the large rapidity region, thus we employ the rapidity cut on the emitted photon as $|y^{\gamma}|<3$.

\begin{table}
\centering
\caption{LDMEs $\langle \mathcal O^{J/\psi(\Upsilon)}[n])\rangle$ used in our calculation, in units of $\textrm{GeV}^3$.}
\label{LDME}
\begin{tabular}{lccccc}
\hline
&  $n=^3S_1^1$  & $n=^1S_0^8$ & $n=^3S_1^8$ & $n=^3P_J^8$    \\ \hline
$J/\psi$(\cite{Zhang:2014ybe,Sun:2015pia})    & 0.645 & $0.78 \times 10^{-2}$ & $1.0 \times 10^{-2}$ & $1.7 \times 10^{-2}$\\
$J/\psi$(\cite{Bodwin:2014gia})    & 1.16 & $9.9 \times 10^{-2}$ & $1.1 \times 10^{-2}$ & $0.49 \times 10^{-2}$\\
$\Upsilon$(\cite{Feng:2015wka})  & 9.28 & $13.6 \times 10^{-2}$ & $0.61 \times 10^{-2}$ & $-0.93 \times 10^{-2}$\\ \hline
\end{tabular}
\end{table}

\subsection{Phenomenological results for $J/\psi$}

\begin{figure}
\includegraphics[width=0.23\textwidth]{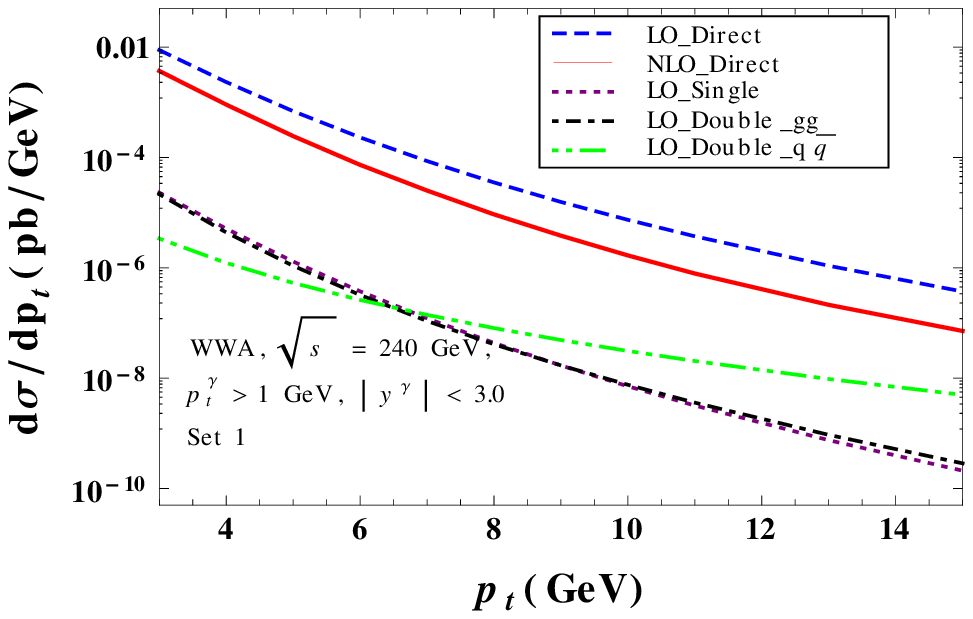}
\includegraphics[width=0.23\textwidth]{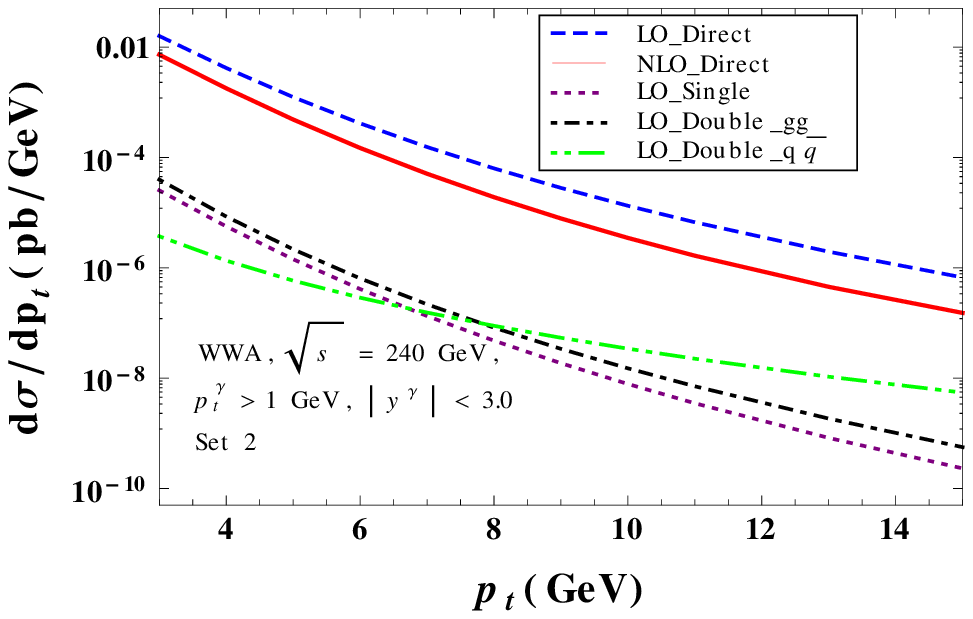}\\
\includegraphics[width=0.23\textwidth]{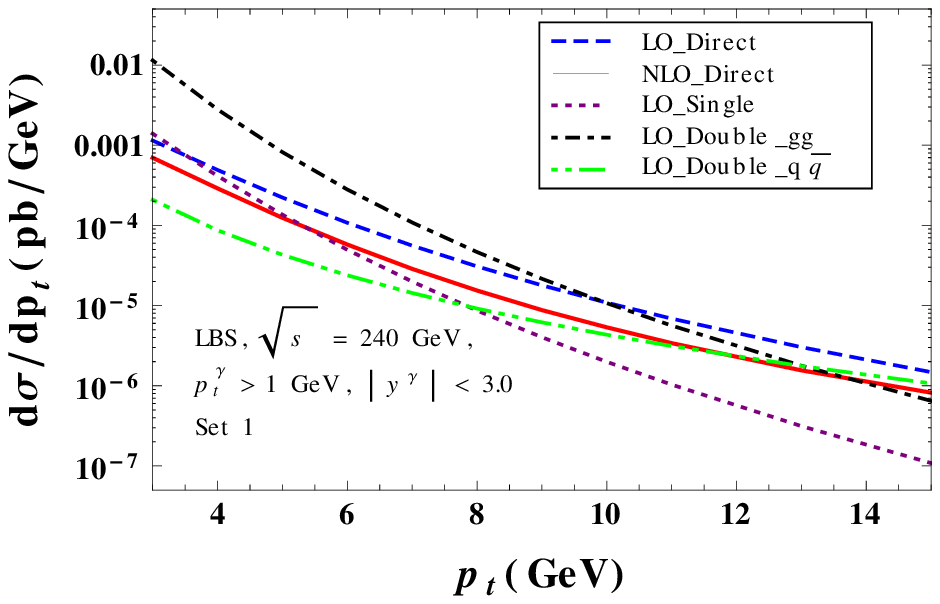}
\includegraphics[width=0.23\textwidth]{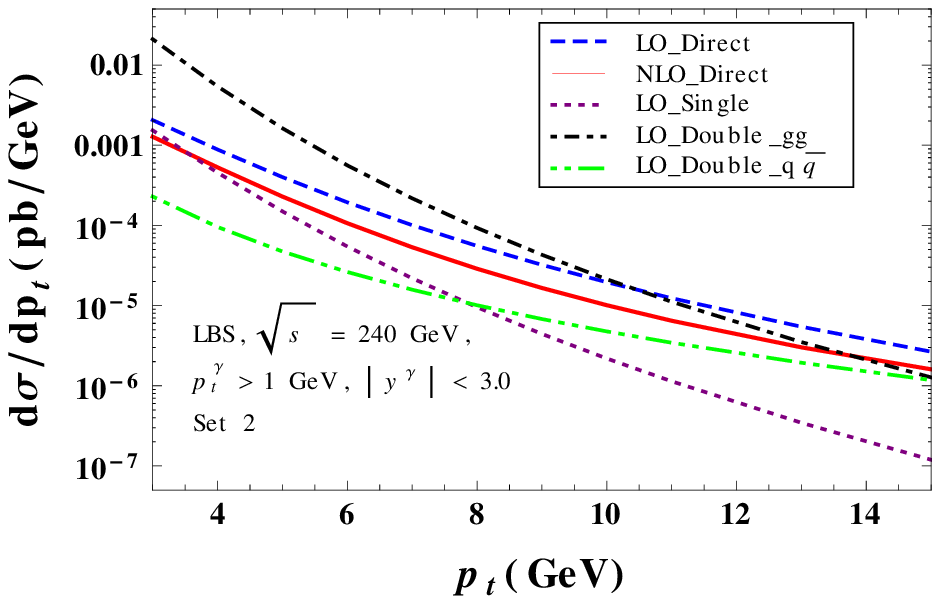}\\
\caption{\label{fig:ptJpsiresolve}
The differential cross sections for the $J/\psi$ production with respect to $p_t$, including the direct, single- and double- resolved processes. The upper two figures are for WWA and the lower two are for LBS. For the resolved processes, the photon PDF from GRS \cite{Gluck:1999ub} is employed.}
\end{figure}

\begin{figure}
\includegraphics[width=0.23\textwidth]{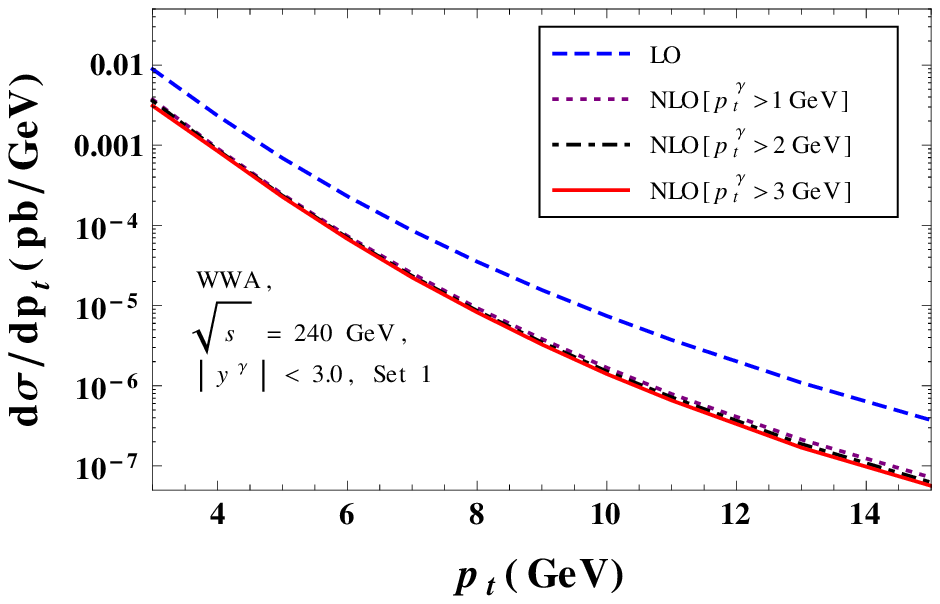}
\includegraphics[width=0.23\textwidth]{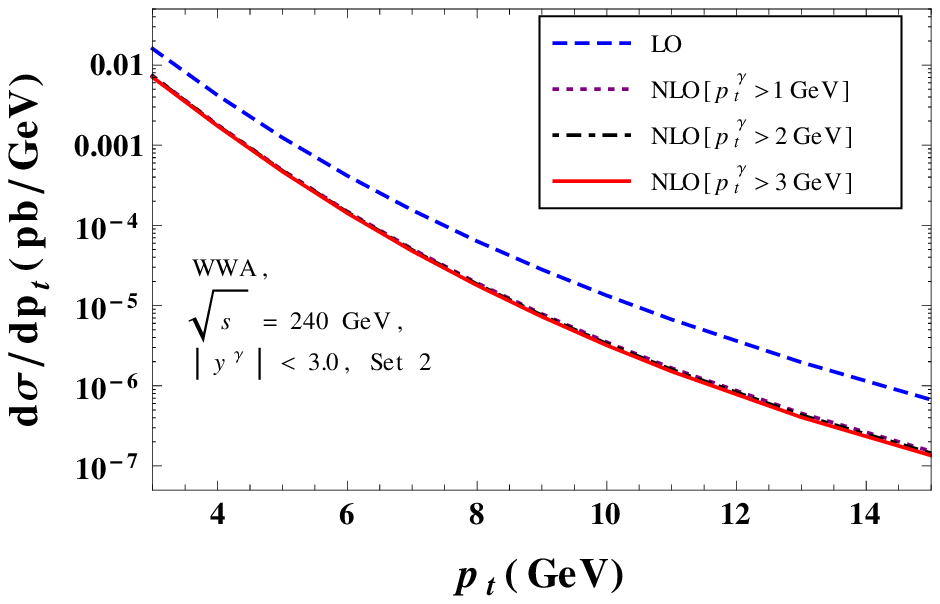}
\caption{\label{fig:ptJpsi}
The differential cross sections for the $J/\psi$ production with respect to $p_t$, applying different $p^{\gamma}_t$ cuts on the final emitted photon. }
\end{figure}

\begin{figure}
\includegraphics[width=0.23\textwidth]{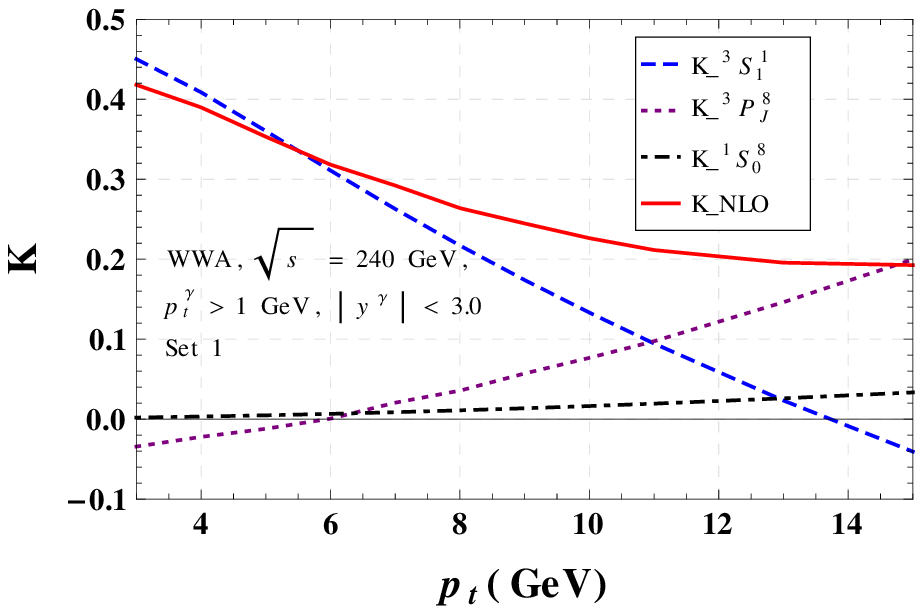}
\includegraphics[width=0.23\textwidth]{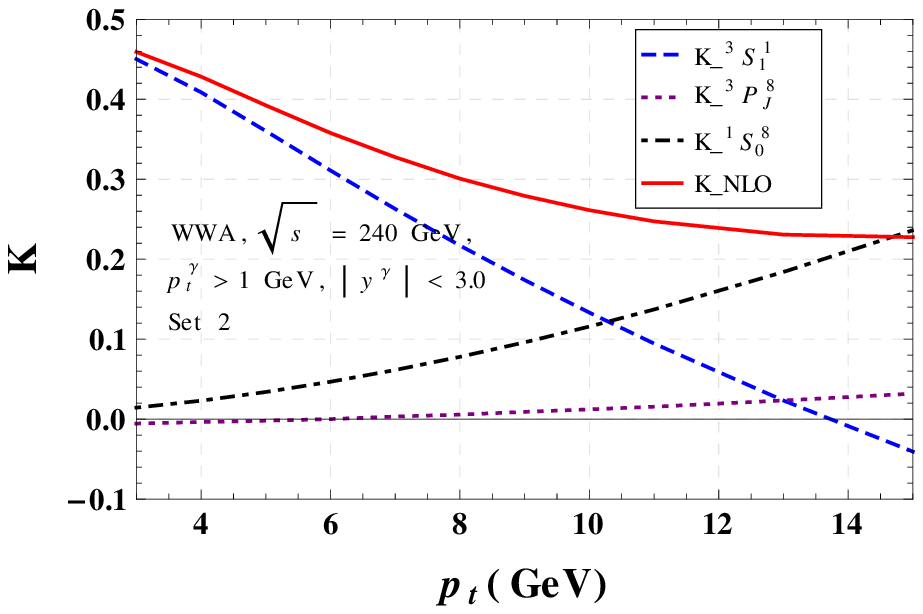}
\caption{\label{fig:KJpsi}
$K$ factor with respect to $p_t$ for different partonic processes.
}
\end{figure}

\begin{figure}
\includegraphics[width=0.3\textwidth]{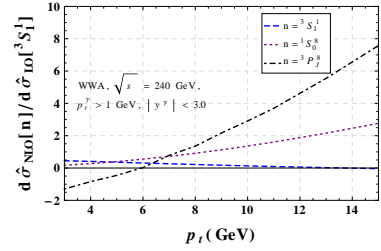}
\caption{\label{fig:SDCKJpsi}
Ratios $d\hat{\sigma}_{NLO}[n]/d\hat{\sigma}_{LO}[^3S_1^1]$ with $n=^3S_1^1,^1S_0^8,^3P_J^8$ as a function of $p_t$ for different partonic processes.
}
\end{figure}

In the first place, we demonstrate the relative significance of the direct, single- and double- photoproductions in Figure \ref{fig:ptJpsiresolve}, including both WWA and LBS. One can clearly see that, for the case of WWA, the contributions via the resolved processes are at least 2 orders of magnitude smaller than the direct one by reason of the mentioned before suppressions of the CO LDMEs and the PDF of the hadronic contents in the photon. As for the case of LBS, the contributions from direct photoproduction are minor while the resolved processes are dominant. At the circular $e^+e^-$ collider CEPC, the initial photons are generated from bremsstrahlung described by the density function of WWA, thus the resolved processes can be safely neglected.

The $p_t$ distributions of $J/\psi$ with different $p^{\gamma}_t$ cuts are presented in Figure \ref{fig:ptJpsi}. As is shown in this figure, the NLO QCD corrections significantly reduce the LO results, and this ``significance" increases with the value of $p_t$. Different $p^{\gamma}_{t,min}$ just slightly impact on the $p_t$ distributions of $J/\psi$. Therefore, without loss of generality, we will only focus our following analysis on this condition $p^{\gamma}_t>1$ GeV. In order to investigate the relative importance of the contributions of these partonic processes listed in Eqs. (4-6), we present the $K-$factors in Figure \ref{fig:KJpsi}. One can find that the virtual QCD corrections to $\gamma\gamma \to J/\psi+\gamma$ is negative, even bigger than the LO cross section when $p_t$ is relatively high, such as 13.5 GeV. As for the CO processes, the ratio of contributions to the total NLO results are continuous increasing with the value of $p_t$. In other words, as the transverse momentum of $J/\psi$ increases, the significant reduction via the virtual corrections will make the effects of the CO processes more and more remarkable. To be specific, in the left figure, the contribution of $^3P_J^8$ will be in the excess of that of $^3S_1^1$ when $p_t>11$ GeV and absolutely dominant when $p_t$ is close to 15 GeV. On account of the relatively small value of $\langle \mathcal O^{J/\psi}[^1S_0^8])\rangle$, the $^1S_0^8$ just provides slight contributions. However in the right figure where the set of LDMEs based on the ``$^1S_0^8$ dominance" are employed, instead of $^3P_J^8$, the $^1S_0^8$ is the dominant one at mid $p_t$ regions. In Figure \ref{fig:SDCKJpsi}, we present the ratio of the SDCs corresponding to different partonic processes listed in (4-6) to the LO SDC, namely $d\hat{\sigma}_{NLO}[n]/d\hat{\sigma}_{LO}[^3S_1^1]$ with $n=^3S_1^1,^1S_0^8,^3P_J^8$. One can find that $d\hat{\sigma}_{NLO}[^3P_J^8]$ rapidly increases with $p_t$, and around $p_t\simeq7 GeV$, the ratio can be bigger than 1. Considering the significant reduction via the virtual corrections, $d\hat{\sigma}_{NLO}[^3P_J^8]$ and $d\hat{\sigma}_{NLO}[^1S_0^8]$ can eventually be orders of magnitude in the excess of $d\hat{\sigma}_{NLO}[^3S_1^1]$. For example,
\begin{eqnarray}
d\hat{\sigma}_{NLO}[^3P_J^8]/d\hat{\sigma}_{LO}[^3S_1^1]|_{p_t=13\textrm{GeV}} &\simeq& 240, \nonumber \\
d\hat{\sigma}_{NLO}[^1S_0^8]/d\hat{\sigma}_{LO}[^3S_1^1]|_{p_t=13\textrm{GeV}} &\simeq& 94.
\end{eqnarray}
So large SDCs are sufficiently able to compensate for the CO LDMEs suppressions.
\begin{figure}
\includegraphics[width=0.23\textwidth]{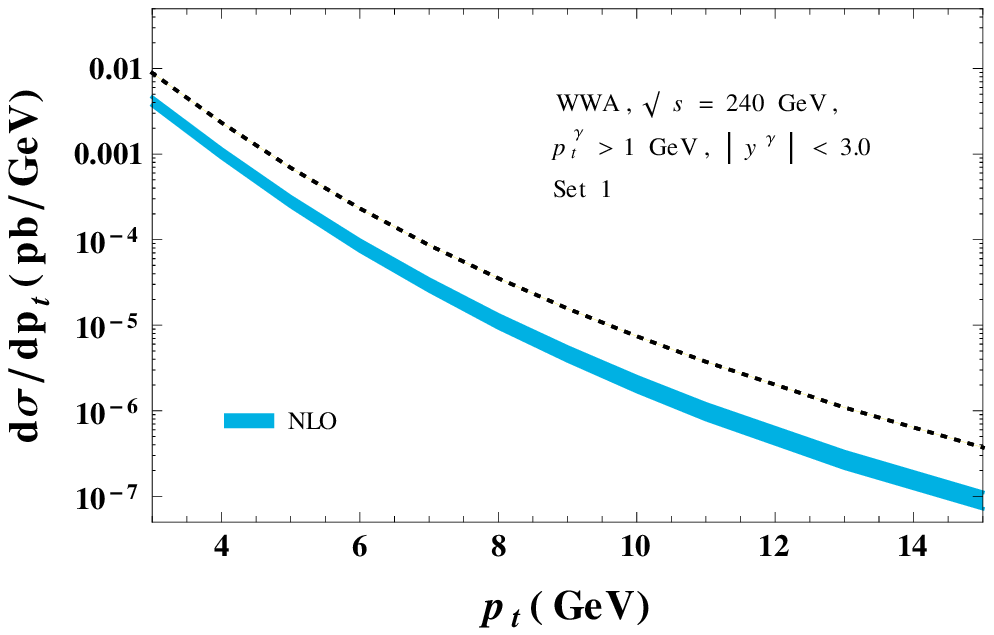}
\includegraphics[width=0.23\textwidth]{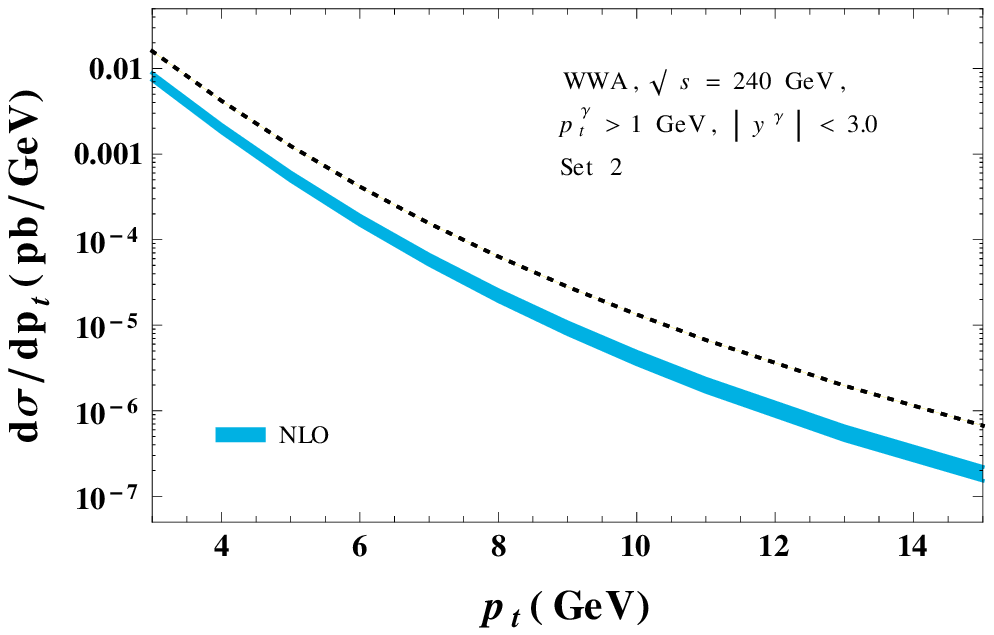}
\caption{\label{fig:ptJpsiuncer}
The differential cross sections for the $J/\psi$ production with respect to $p_t$, applying $p^{\gamma}_t>1$ GeV. The band in these two figures are caused by the variation of $\mu_r$ from $\mu_0$ to $2\mu_0$ and the dotted lines denote the LO results.
}
\end{figure}

\begin{figure}
\includegraphics[width=0.23\textwidth]{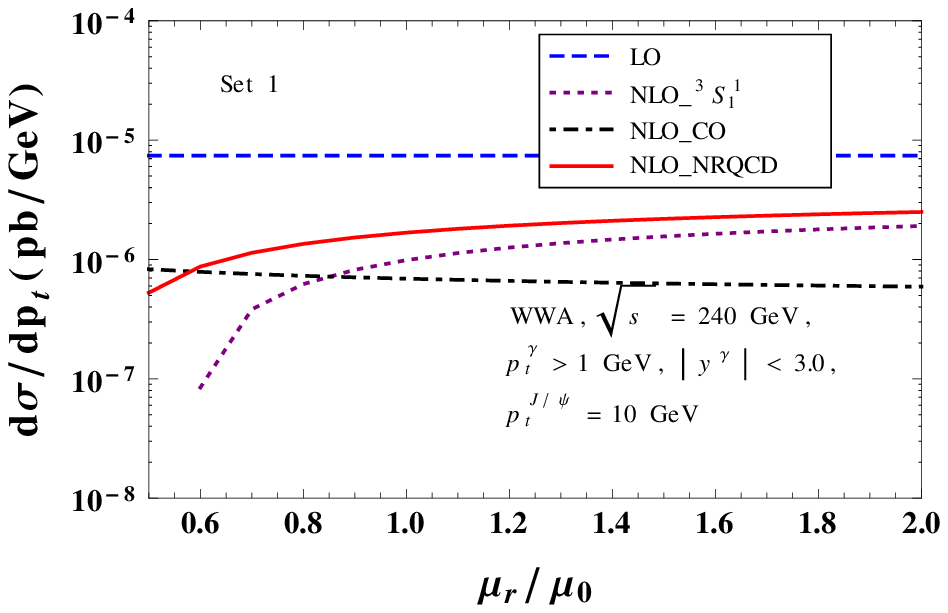}
\includegraphics[width=0.23\textwidth]{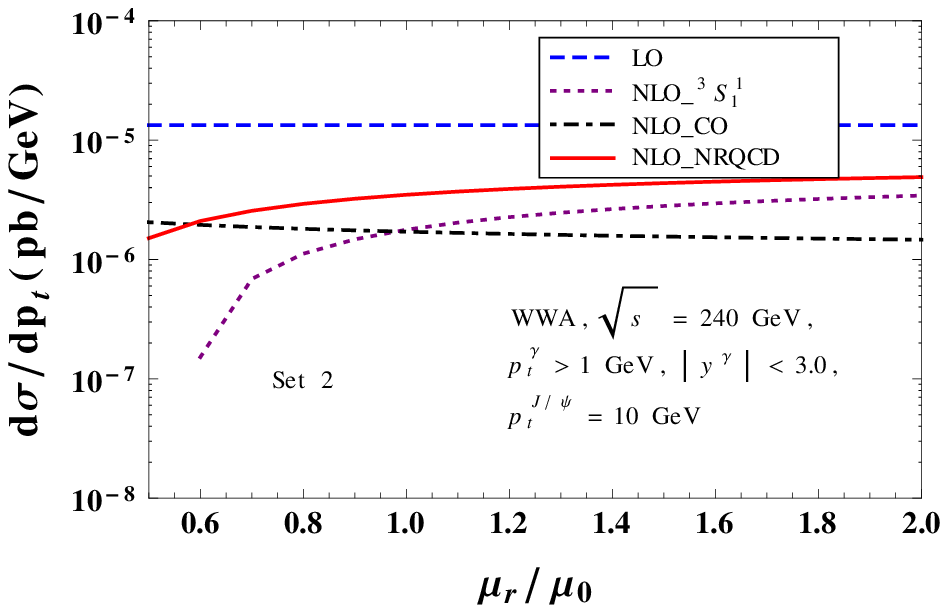}
\caption{\label{fig:Jpsimu}
Differential cross section at $p_t=10$ GeV as a function of the renormalization scale $\mu_r$.
}
\end{figure}
Notice that the LO process $\gamma\gamma \to J/\psi+\gamma$ is a pure QED process, thus the NLO QCD results may suffer heavily from the uncertainty of the renormalization scale $\mu_r$, just like the case of $e^+e^- \to \gamma^{*}\gamma^{*} \to J/\psi+J/\psi$ where the NLO QCD corrections change significantly with the variation of $\mu_r$ \cite{Gong:2008ce}. In Figure \ref{fig:ptJpsiuncer}, one can find that the band reflecting the impact of the uncertainty of $\mu_r$ is narrow, indicating a moderate dependence of the yields on the renormalization scale. To be specific,
\begin{eqnarray}
\left(\frac{d\sigma_{NLO,\mu_r=2\mu_0}}{d\sigma_{NLO,\mu_r=\mu_0}}\right)|_{p_t=7\textrm{GeV}} &=& (1.37_{\textrm{Set} 1},1.31_{\textrm{Set} 2}), \nonumber \\
\left(\frac{d\sigma_{NLO,\mu_r=2\mu_0}}{d\sigma_{NLO,\mu_r=\mu_0}}\right)|_{p_t=10\textrm{GeV}} &=& (1.49_{\textrm{Set} 1},1.40_{\textrm{Set} 2}) \nonumber \\
\end{eqnarray}
Taking $p_t=10$ GeV for example, the dependences of the differential cross sections corresponding to different partonic processes on $\mu_r$ are presented in Figure \ref{fig:Jpsimu}. Since the virtual correction to $^3S_1^1$ is negative, the differential cross section based on color singlet mechanism will increase with the value of $\mu_r$, which is responsible for the eventual monotonic increase behavior of the NRQCD results.

\begin{figure}
\includegraphics[width=0.23\textwidth]{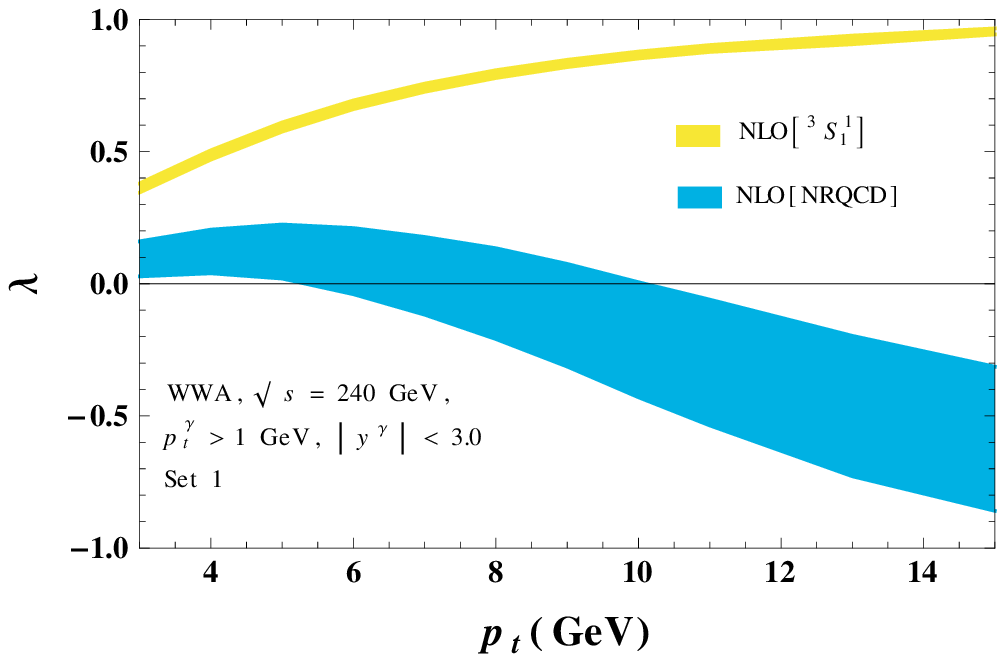}
\includegraphics[width=0.23\textwidth]{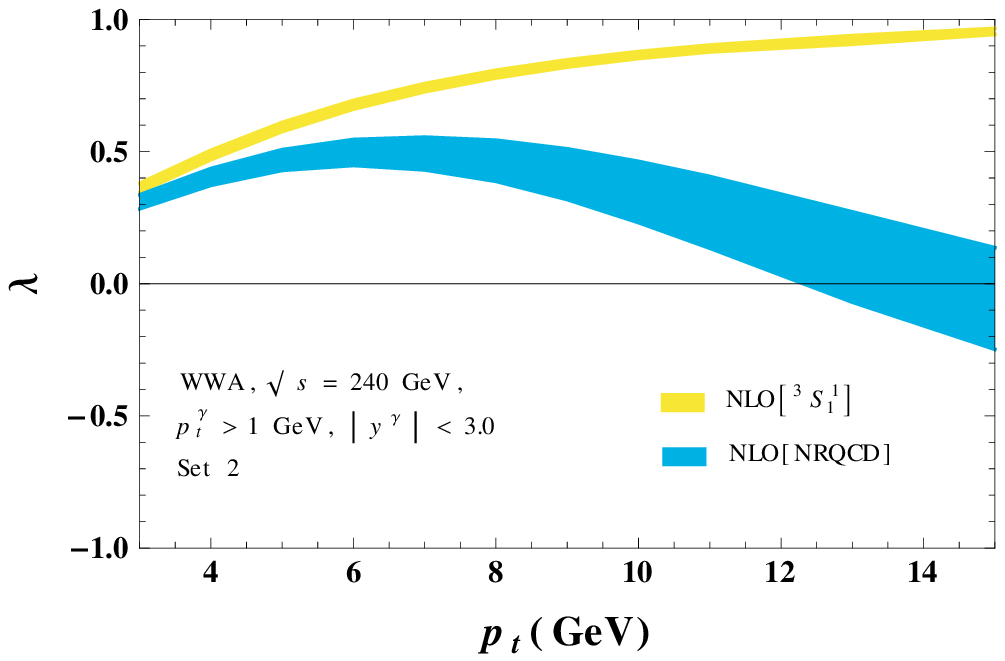}
\caption{\label{fig:polJpsi}
The polarization parameter $\lambda$ with respect to $p_t$, applying $p^{\gamma}_t>1$ GeV. The band in these two figures arise from the variation of $\mu_r$ from $\mu_0$ to $2\mu_0$.
}
\end{figure}

Reviewing the well known polarization puzzle of $J/\psi$, the predicted polarization via NLO NRQCD is slightly transverse polarized, which is far different from the results of CS (totally transverse for LO and longitudinal for NLO). Namely, the inclusion of CO partonic processes may dramatically change the polarizations. Considering the CO contributions in our concerned process is increasingly important with the value of $p_t$, thus it is interesting and natural to investigate the polarization of $\gamma\gamma \to J/\psi+\gamma+X$. In Fiure \ref{fig:polJpsi}, we present the predicted polarization based on CS and NRQCD, respectively. One can see that, the polarization via CS is totally transverse, in contrary to the one based on NRQCD which varies from transverse to longitudinal. Such significant difference can be employed to serve as an ideal laboratory to further identify the significance of the color octet mechanism. The distinct differences between the two predicted polarizations can be attributed to the following aspects: 1) the color octet contributions are significant, especially at mid $p_t$ regions; 2) in Eq. \ref{softhard}, the $^3S_1^1$ content involved in the soft part of the cross section of $^3P_J^8$ provides a large negative transverse contribution and a relatively tiny negative longitudinal one; 3) the polarization of the ``hard" part of $^3P_J^8$ in Eq. \ref{softhard} is slightly longitudinal; 4) the polarization of $^1S_0^8$ is unpolarized. In view of these points, the inclusion of CO contributions will thoroughly change the configuration of the polarization, from transverse ($^3S_1^1$) to longitudinal (NRQCD). Notice that, unlike the case of yields, the variation of $\mu_r$ can have a significant effect on the polarization of $J/\psi$.

\begin{figure}
\includegraphics[width=0.23\textwidth]{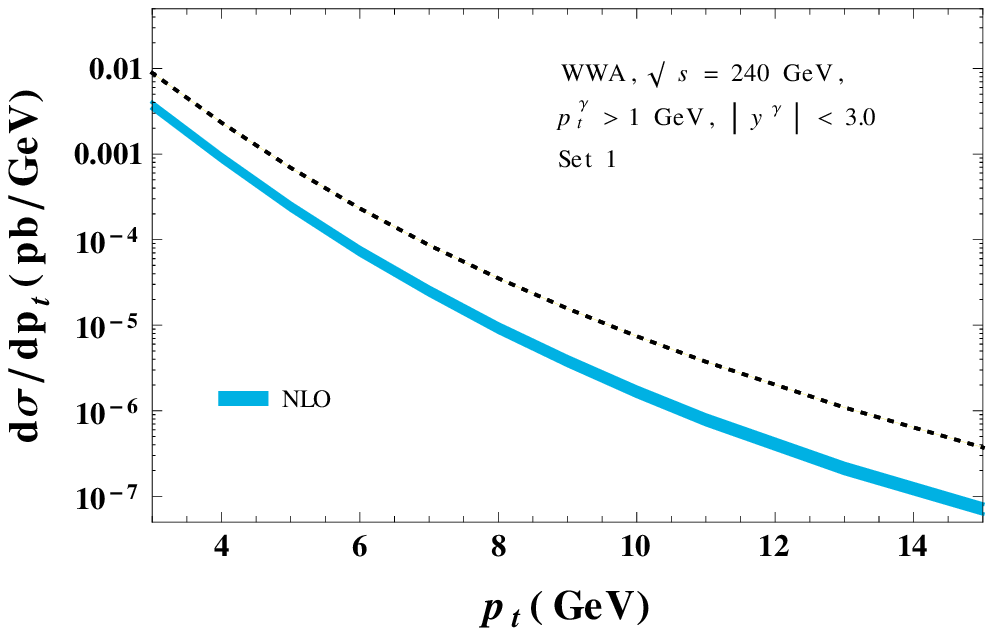}
\includegraphics[width=0.23\textwidth]{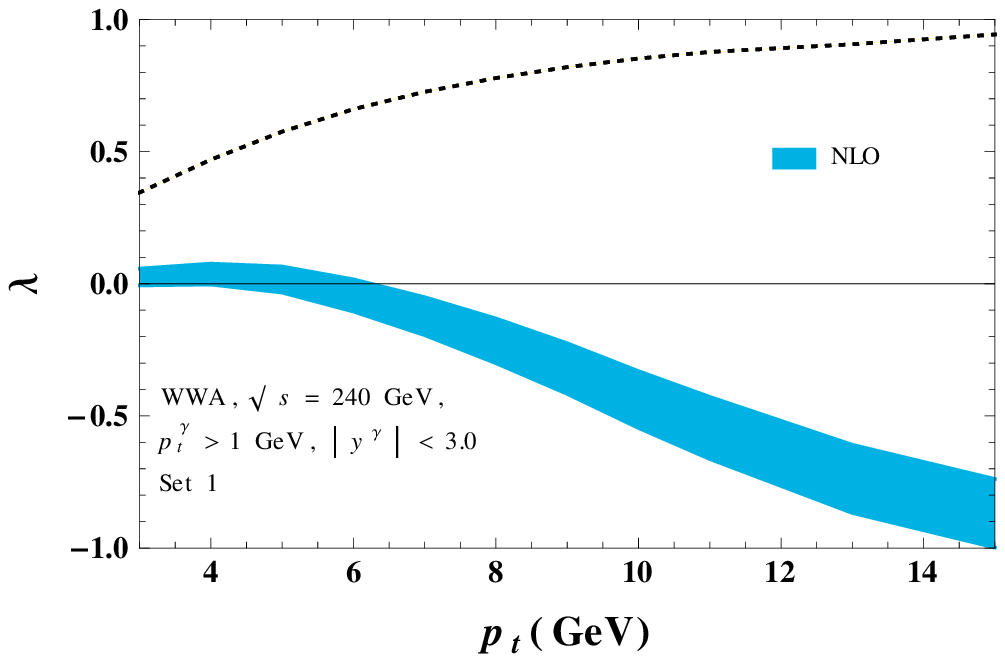}
\caption{\label{fig:ptpolJpsilambda}
The $p_t$ distributions (left) and the polarization parameter $\lambda$ (right) with respect to $p_t$. The bands in these two figures are caused by the variation of $\mu_{\lambda}$ from $0.5m_c$ to $2m_c$ and the dotted lines denote the LO result.
}
\end{figure}
In addition to $\mu_r$, the effects from the uncertainties of $\mu_{\lambda}$ are as well investigated, which are found to be more moderate, as is shown in Figure \ref{fig:ptpolJpsilambda} (Set 1). For the LDMEs of Set 2 based on ``$^1S_0^8$ dominance ", the dominant $^1S_0^8$ partonic process is independent of $\mu_{\lambda}$, thus the variation of $\mu_{\lambda}$ just cause slight effects.

\begin{table}
\centering
\caption{Integrated cross section (unit: \textbf{pb}) for $\gamma\gamma \to J/\psi+\gamma+X$ with $p^{J/\psi,\gamma}_t>1$ GeV and $|y^{\gamma}|<3$. $m_c=1.5$ GeV.}
\label{sigma}
\begin{tabular}{lccccc}
\hline
Set 1 &  \textrm{LO}  & $\textrm{NLO}[^3S_1^1]$ & $\textrm{NLO}$ & $K$    \\ \hline
$\mu_r=\mu_0$    & 0.100 & $0.048$ & $0.042$ & $0.42$\\
$\mu_r=2\mu_0$    & 0.100 & $0.058$ & $0.053$ & $0.53$\\ \hline
Set 2 &  \textrm{LO}  & $\textrm{NLO}[^3S_1^1]$ & $\textrm{NLO}$ & $K$    \\ \hline
$\mu_r=\mu_0$    & 0.180 & $0.086$ & $0.085$ & $0.48$\\
$\mu_r=2\mu_0$    & 0.180 & $0.104$ & $0.103$ & $0.58$\\ \hline
\end{tabular}
\end{table}

In Table \ref{sigma}, we present the integrated cross section with two renormalization scales, imposing two cuts on the transverse momentum of $J/\psi$ and the emitted photon, namely $p^{J/\psi,\gamma}_t>1$ GeV. Meanwhile the rapidity of the final photon is limited as $|y^{\gamma}|<3$. One can find that, for the integrated cross section, the CS contribution is dominant over the CO ones, which are minor at low $p_t$ regions. The variations of $\mu_r$ from $\mu_0$ to $2\mu_0$ just change the cross section by about $26\%$ for Set 1 and $21\%$ for Set 2, indicating a moderate dependence on the renormalization scale.

\subsection{Phenomenological results for $\Upsilon$}

\begin{figure}
\includegraphics[width=0.23\textwidth]{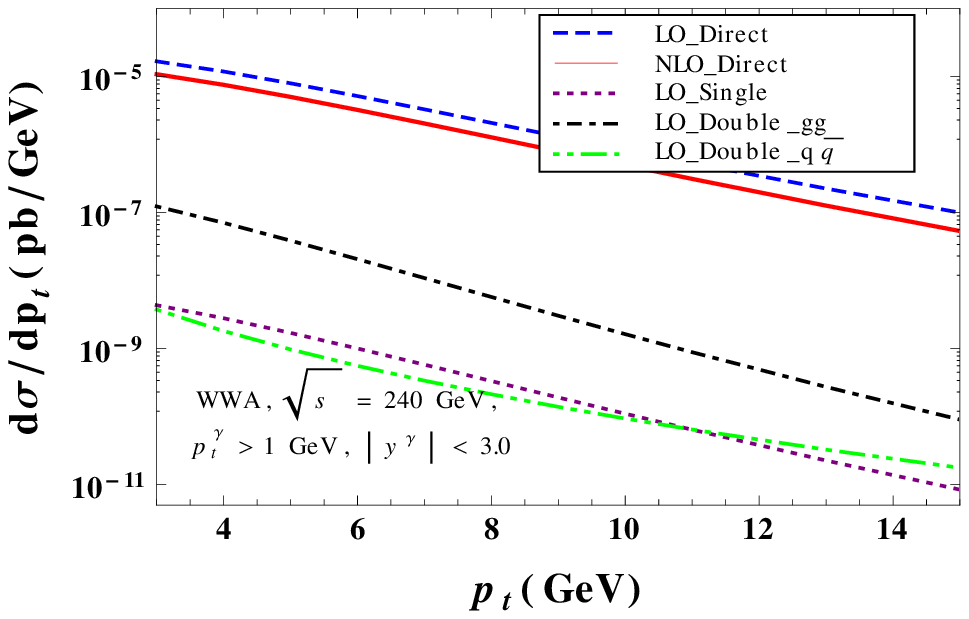}
\includegraphics[width=0.23\textwidth]{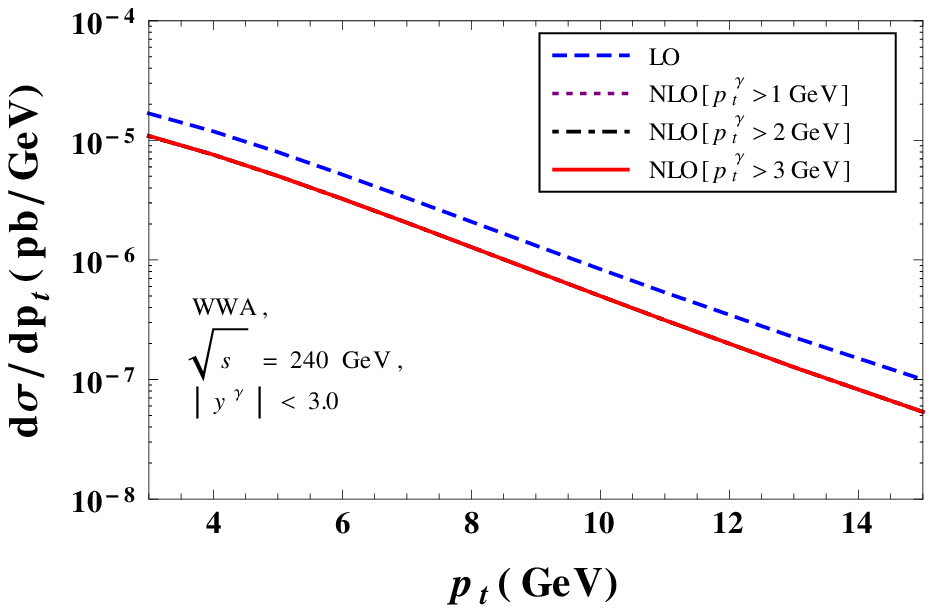}
\caption{\label{fig:ptupsilon}
The left figure represents the differential cross sections for the $\Upsilon$ production with respect to $p_t$, including both direct and resolved processes. The right one denotes the $p_t$ distributions of $\Upsilon$ via the direct photoproduction $\gamma \gamma \to \Upsilon+\gamma+X$, imposing different $p_t^{\gamma}$ cuts on the emitted photon.
}
\end{figure}

\begin{figure}
\includegraphics[width=0.23\textwidth]{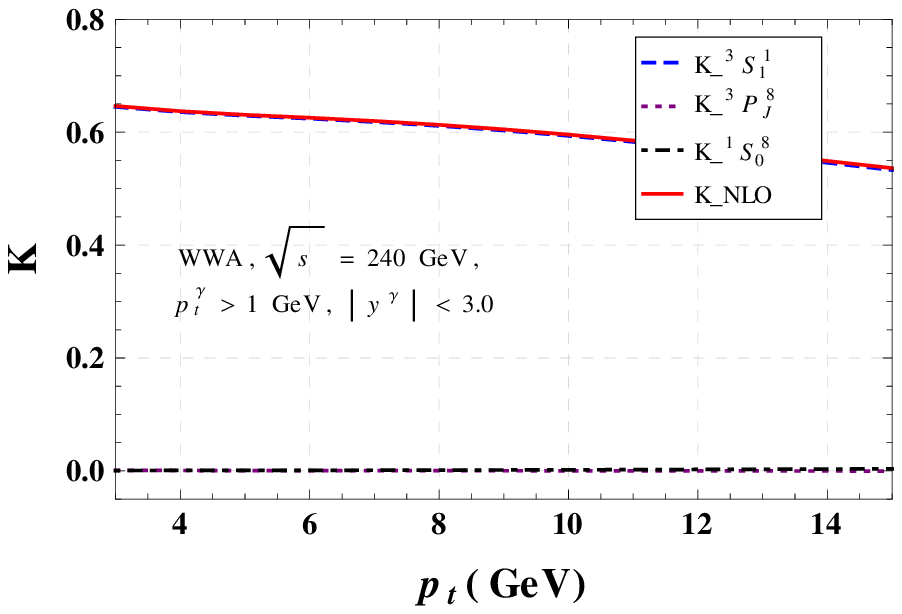}
\includegraphics[width=0.23\textwidth]{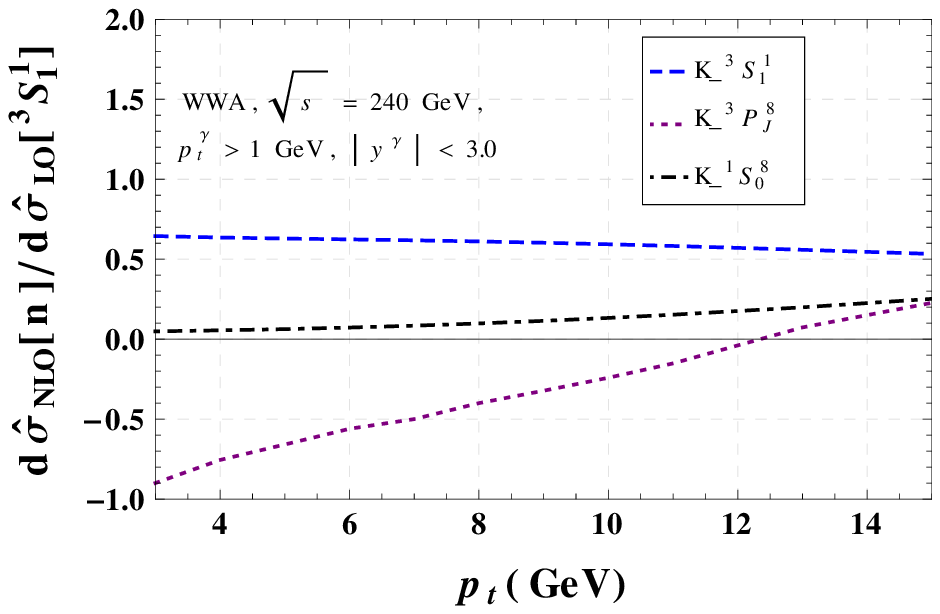}
\caption{\label{fig:Kupsilon}
The left figure is for the $K$ factor with respect to $p_t$ for different partonic processes. And the right figure is for the ratio $d\hat{\sigma}_{NLO}[n]/d\hat{\sigma}_{LO}[^3S_1^1]$ with $n=^3S_1^1,^1S_0^8,^3P_J^8$.
}
\end{figure}

\begin{figure}
\includegraphics[width=0.23\textwidth]{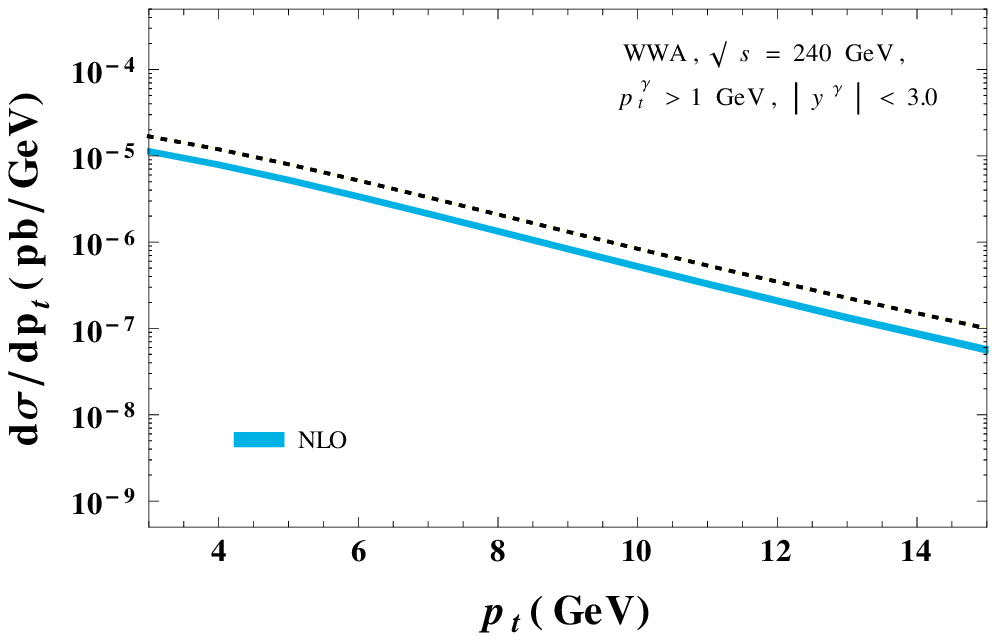}
\includegraphics[width=0.23\textwidth]{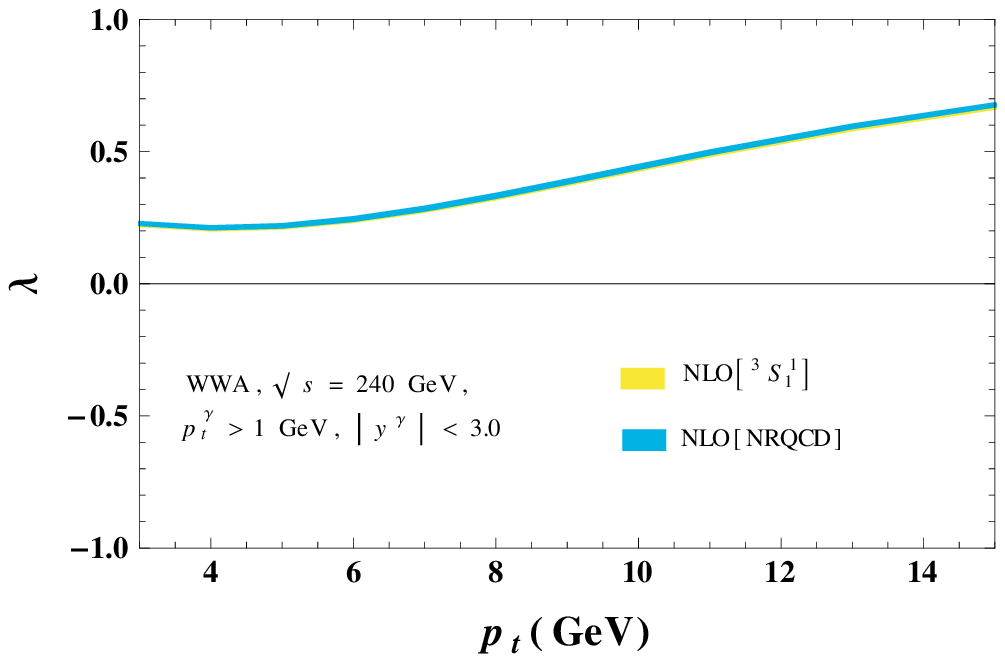}
\caption{\label{fig:ptpolUpsilon}
The $p_t$ distributions for $\Upsilon$ (left) and the polarization parameter $\lambda$ (right) with respect to $p_t$. The bands in these two figures are caused by the variation of $\mu_r$ from $\mu_0$ to $2\mu_0$ and the dotted line in the left figure denotes the LO result. Notice that, since the CO contributions are negligible, in the right figure, the bands for the CS and NRQCD results almost overlap each other.
}
\end{figure}

Now we are in a position to present the phenomenological results for the production of $\Upsilon$. From the left figure of Figure \ref{fig:ptupsilon}, like the case of $J/\psi$, the direct photoproductions play the most important role and the resolved processes can be safely neglected. As is shown in the right one, the three lines for different $p^{\gamma}_t$ cuts overlap each other, implying the contributions from CO processes are tiny. For the sake of further investigating the relative significance of different partonic processes, we present the $K-$factor and the ratio $d\hat{\sigma}_{NLO}[n]/d\hat{\sigma}_{LO}[^3S_1^1]$ with $n=^3S_1^1,^1S_0^8,^3P_J^8$ in Figure \ref{fig:Kupsilon}. From the right one, we find that, unlike the case of $J/\psi$, the CO SDCs are just the same order of the color singlet one, hence the significant suppression from the CO LDMEs will make the color octet contributions almost negligible, as is shown in the left figure. Comparing with the $K$-factor of $J/\psi$ in Figure \ref{fig:KJpsi}, as expected, the convergence of the perturbative calculations for $\Upsilon$ is indeed much better than that of $J/\psi$. As for the polarization parameter, the predicted NLO results are totally dominated by the color singlet contributions, as is shown in the figure on the right hand side of Figure \ref{fig:ptpolUpsilon}. Meanwhile, from the left figure, the dependence on the renormalization scale $\mu_r$ is found to be quite moderate through the narrow bands arisen from the variation of $\mu_r$ from $\mu_0$ to $2\mu_0$. In addition, we find that the uncertain of $\mu_{\lambda}$ just bring about an insignificant effect.

\section{Summary} \label{sec:summary}
In this paper, based on the factorization of the non-relativistic QCD, we systematically investigate the semi-inclusive photoproduction of $J/\psi$ and $\Upsilon$ via $\gamma\gamma \to J/\psi(\Upsilon)+\gamma+X$ at CEPC up to $\mathcal O(\alpha^3\alpha_s)$, including both yields and polarizations. For the color octet contributions, two different sets of LDMEs are employed to provide the NRQCD predictions for $J/\psi$ and one set for $\Upsilon$. We find that the NLO corrections will significantly reduce the LO results which can be attributed to that the virtual corrections to $^3S_1^1$ is large and negative. For $J/\psi$ production, the color octet contributions are increasingly significant with the values of $p_t$, by which the polarizations of $J/\psi$ can be changed dramatically from toally transverse to longitudinal. This difference can be regarded as a distinct signal for identifying the significance of the CO mechanism. While, for the case of $\Upsilon$, the effects of the color octet processes is negligible, both for yields and polarizations. The future measurements on this semi-inclusive photoproduction of $J/\psi(\Upsilon)+\gamma+X$ in photon-photon collisions, especially on the polarization parameters of $J/\psi$, will be a ideal laboratory for the study of heavy quarkonium production mechanism and helpful to clarify the problems of the $J/\psi$ polarization puzzle.

\hspace{1cm}

\noindent{\bf Acknowledgments}:
We thank Dr. Hong Fei Zhang for helpful discussions. This work are supported by the Natural Science Foundation of China under Grants No. 11647112 and No. 11647113.\\

\end{document}